\documentclass[aps,pra,multicolumn,twocolumn]{revtex4}
\usepackage{graphicx}
\usepackage{mathrsfs}
\usepackage{amssymb}
\usepackage{amsmath}

\begin{document}

\title{Quantum gates  by periodic driving}

\author{Z. C. Shi$^{1,2}$ and X. X. Yi$^{1}$\footnote{Corresponding  address:
yixx@nenu.edu.cn}}
\affiliation{
$^1$ Center for Quantum Sciences and School of Physics, Northeast
Normal University, Changchun 130024, China\\
$^2$ School of Physics and Optoelectronic Technology\\
Dalian University of Technology, Dalian 116024 China}

\date{\today}

\begin{abstract}
Topological quantum computation has been extensively  studied
due to its robustness against decoherence. A conventional
way to realize it is by adiabatic operations---it
requires relatively long time to accomplish so that the speed of quantum computation slows down. In
this work, we present a method to realize
topological quantum computation by periodic driving. Compared to the
adiabatic evolution, the total operation time can be regulated arbitrarily
by the amplitude and frequency of the periodic driving.
For the sinusoidal driving, we give an expression for
the total operation time in the high-frequency limit.
For the square wave driving, we derive an exact
analytical expression for the evolution operator without any approximations, and show that the amplitude and
frequency of driving field depend on its period and total operation
time. This could provide a new direction in regulations of the
operation time in topological quantum computation.
\end{abstract}

 \maketitle

\section{introduction}

Decoherence is an enemy of quantum computation, which is the loss of
coherence  due to the presence of environments. As a promising
avenue to deal with the decoherence, topological quantum
computations
\cite{kitaev03,bonderson08,nayak08,miyake10,akhmerov10,xue14a,MONG14,wootton14}
employ two-dimensional quasiparticles called anyons, whose world
lines cross over one another to form braids in a three-dimensional
spacetime. Information encoded in the anyons is robust against local
perturbations and quantum operations can be performed by braiding
the non-Abelian anyons \cite{kitaev03,kitaev06}. The simplest example
of the non-Abelian anyons is the Majorana fermions which are
predicted to exist in  fractional quantum Hall systems
\cite{read00},  topological insulators \cite{hasan10,qi11}, solid
state systems \cite{alicea12},  and   semiconductor-superconductor
hybrid systems \cite{sau2010,oreg10,lutchyn10}.  The signatures of
Majorana fermions have also been observed in experiments more
recently \cite{das2012,mourik12,rokhinoson12,perge14,lee14}, which gives rise to
an opportunity  to encode a qubit  by Majorana fermions in these
materials.

A quantum task is  often accomplished by a sequence  of quantum
operations rather than single quantum operation
\cite{sau11,heck12,kraus13,laflamme14,ckarje11,chiu12,alicea11}.
The total operation time increases linearly with the
increasing of the number of operations. Considering the limit of the
coherence time of quantum systems, long operation time is not
favorable, even if the quantum topological
computation is robust against perturbations. On the other hand, the
time-periodic driving systems have been extensively studied in the
past few years. Especially, several work
\cite{oka09,kitagawa10,linder11,gu11,jiang11,liu13,katan13,rudner13,torres14,benito14,zhou14,piskunow14,usaj14}
have shown that the topological properties can be changed in
topologically trivial system by time-periodic driving (e.g., the
existence of Floquet topological insulators or Floquet Majorana
fermions).
Recently, the Floquet Majorana
fermions is realized by periodic driving fields in the system of coupled quantum
dots proximity to a $s$-wave superconductor \cite{li2014}.
More recently, it has been proposed to achieve the direct
coupling between the topological and conventional qubits by periodic
driving fields \cite{xue14b}. In this paper we explore the
possibility  to regulate the total operation time by periodic driving
For concreteness, the physical model of interest is the
quantum dots coupled to the Majorana modes in a topological
superconductor. Of course, this method can also be extended to the
other quantum systems.

The paper is organized as follow. In Sect. \ref{II},  we briefly
introduce the topological quantum computation by   adiabatic
evolution. In Sect. \ref{III}, we first recall the Floquet theory,
then the periodic driving fields in the form of   sinusoidal, square
wave, and $\delta$-function kick  are applied separately  to
modulate the total operation time for realizing the quantum
operations. Finally we extend this method to other hybrid quantum
systems in Sect. \ref{V}. The discussion and conclusion are given in Sect. \ref{IV}.

\section{quantum computation by adiabatical evolution} \label{II}

Recently, the adiabatic evolution has widely  applied to the preparation and
manipulation of Majorana fermions \cite{ckarje11,chiu12,kraus12}. In particular, it
has been shown that topological quantum information processing
becomes possible in the one-dimensional network \cite{alicea11} by
adiabatically controlling the locally tunable gates which affect the chemical
potential over a finite length of the wire.
In following we describe the main idea of adiabatic
evolution. That is, design a Hamiltonian $H_1$ whose ground state is the
target state $|\Psi_T\rangle$ while the ground state
$|\Psi_0\rangle$ of Hamiltonian $H_0$ is easily to prepared. Assume
that there exists a quantum system satisfying the following Hamiltonian
\begin{eqnarray}\label{1}
H=[1-f(\frac{t}{T})]H_0+f(\frac{t}{T})H_1,
\end{eqnarray}
where $f(t)$ is a slowly varying function of evolution
time $t$ with $f(0)=0$ and $f(1)=1$. According to the adiabatic
theorem, the quantum system evolves
adiabatically from the initial (ground) state $|\Psi_0\rangle$
to the target (ground) state $|\Psi_T\rangle$ at time $t=T$.

\begin{figure}[h]
\centering
\includegraphics[scale=0.5]{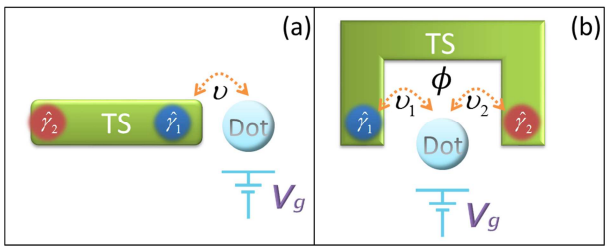}
\caption{The setup for realizing the operation $P_1$.} \label{fig:01}
\end{figure}

In the present work, the physical model of interest consists of a
quantum dot coupled  to a semiconducting nanowire, as shown in  Fig.
\ref{fig:01}(a). In a magnetic field, by proper spin orbit interaction
and proximity coupling to a superconductor, the nanowire can exist
the Majorana bound states in the topological phase
\cite{lutchyn10,oreg10,alicea10,stoudenmire11}. Then the effective
Hamiltonian (in the low-energy limit) for the quantum dot coupling
to the Majorana mode reads \cite{flensberg11}
\begin{eqnarray}   \label{2}
H=\varepsilon(t) \hat{a}^{\dag}\hat{a}+(v^{\ast}\hat{a}^{\dag}-v\hat{a})\hat{\gamma}_{1},
\end{eqnarray}
where $a$ ($a^{\dag}$) is the annihilation (creation)  operator for
the electron in quantum dot and the on-site energy $\varepsilon(t)$
for the quantum dot can be controlled by the gate voltage $V_g$. $v$
denotes the tunnel coupling between the quantum dot and the Majorana
mode $\hat{\gamma}_{1}$. Without loss of generality we assume $v$ is
an real number and take all physical parameters in units of $v$.
Since the Majorana mode $\hat{\gamma}_{i}$ is Hermitian
($\hat{\gamma}^{\dag}_{i}=\hat{\gamma}_{i}$ and
$\hat{\gamma}^2_{i}=1$), we cannot use the number operator
$\hat{\gamma}^{\dag}_{1}\hat{\gamma}_{1}$ to count the occupation
of the Majorana mode. Whereas, two Majorana modes can be
combined to generate one ordinary fermion, e.g.,
$\hat{\gamma}_{1}=\hat{b}+\hat{b}^{\dag}$ and
$\hat{\gamma}_{2}=i(\hat{b}^{\dag}-\hat{b})$. One can
adopt the number operator $b^{\dag}b$ of the ordinary fermion to
count the Majorana modes.

Since the total parity of the electron in quantum dot and  the ordinary
fermion formed by Majorana modes is conserved, the Hamiltonian is
block diagonal in the basis spanned by
$\{|0\rangle_F|0\rangle_D,|1\rangle_F|1\rangle_D,|1\rangle_F|0\rangle_D,|0\rangle_F|1\rangle_D\}$,
\begin{eqnarray}
H=\left(
    \begin{array}{cccc}
      0 & v & 0 & 0 \\
      v & \varepsilon(t) & 0 & 0 \\
      0 & 0 & 0 & v \\
      0 & 0 & v & \varepsilon(t) \\
    \end{array}
  \right),
\end{eqnarray}
where the state $|m\rangle_F|n\rangle_D$ ($m,n=0,1$)  represents $m$
ordinary fermion formed by Majorana mode and $n$ electron in the
quantum dot. In Ref. \cite{flensberg11}, it suggests that by
adiabatically changing the values of $\frac{\varepsilon}{v}$ from
$-\infty$ to $+\infty$, it can realize the operation $P_1$ which
denotes the inversion of the occupation in ordinary fermion combined
by the Majorana mode (i.e., $P_1=\hat{\gamma}_1$),
\begin{eqnarray}
P_1(\sin\theta|0\rangle_F+\cos\theta|1\rangle_F)=\sin\theta|1\rangle_F+\cos\theta|0\rangle_F.
\end{eqnarray}

\begin{figure}[h]
\centering
\includegraphics[scale=0.45]{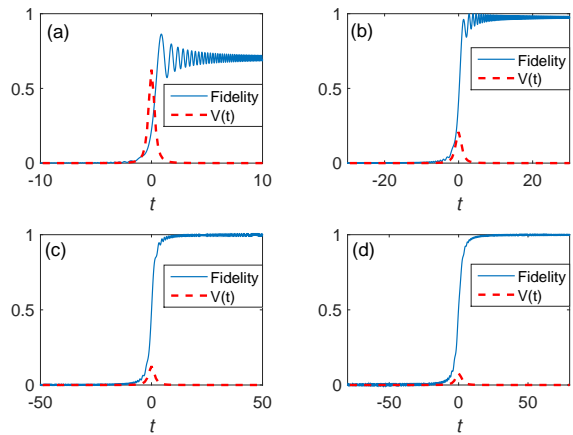}
\caption{The time evolution of the system to obtain the  operation
$P_1$, where the fidelity is defined as
$|\langle\psi(t)|\psi(T)\rangle|^2$. The expression of $V(t)$ is
defined as, $V(t)=|\frac{\langle
E_1(t)|\dot{E}_2(t)\rangle}{E_1(t)-E_2(t)}|$ where $|E_i(t)\rangle
(i=1,2)$ is the instantaneous eigenstate with corresponding
eigenvalue $E_i(t)$ in the even (or odd) subspace (The adiabatic
condition can be written as $V(t)\ll1$).
 The initial state is $|\psi(0)\rangle=\sin\theta|0\rangle_F
+\cos\theta|1\rangle_F$ and the target state is
$|\psi(T)\rangle=\sin\theta|1\rangle_F+\cos\theta|0\rangle_F$,
$\theta=\frac{\pi}{6}$. We have set the on-site energy of the
quantum dot increases with time linearly, e.g.,
$\varepsilon(t)=\frac{50}{T}t$. The operation time is $2T$ and the
final value of $\varepsilon(t)$ is 50 during the time evolution.
All parameters are in units of the tunnel coupling
$v$. (a) $T=10$. (b) $T=30$. (c) $T=50$. (d) $T=80$. The larger $T$ means the smaller of the changing rate
$\frac{\varepsilon}{v}$.
One can find that the operation time for perfectly achieving the
operation $P_1$ is about $2T=100$. } \label{fig:02}
\end{figure}

Fig. \ref{fig:02} shows the different dynamics behaviors  for
distinct changing rate $\frac{\varepsilon}{v}$. It can be
observed in Fig. \ref{fig:02}(a) that the operation $P_1$ cannot be
achieved perfectly since the changing of $\frac{\varepsilon}{v}$
does not satisfy the adiabatic condition very well (It cannot
satisfy $V(t)\ll1$ all the time). Thus the changing rate
$\frac{\varepsilon}{v}$ should be small in order to meet the
adiabatic condition, along with the increasing of operation time
(cf. Fig. \ref{fig:02}(c)-(d)).
To implement the single qubit rotation or non-Abelian operation,
one shall successively execute the operation $P_1$.
Therefore the total operation time increases with the increasing of the
number of operation $P_1$. In addition, it needs to point out that
this Majorana based qubits may be susceptible to decoherence due to the
electron tunnel coupling process \cite{budich12,goldstein11,rainis12,mazza13}. As
a consequence, the adiabatic evolution is at a disadvantage in minimizing the influence of
decoherence as far as possible. Recently, it is overcome by shortcuts to
adiabaticity for the non-Abelian braiding with Y-junction structure \cite{karzig15}. In following it demonstrates
that the situation can be changed by periodic driving.

\section{quantum gates with periodic driving} \label{III}

\subsection{Floquet theory}

Let us first recall the Floquet theory briefly \cite{shirley65}. Provided that the
system Hamiltonian has a time-periodic driving field, $H(t)=H(T+t)$,
where $T$ is the period and the driving frequency reads
$\omega=\frac{2\pi}{T}$. The Floquet theory asserts that the
solutions of Schr\"odinger equation have the form
$|\Phi_n(t)\rangle=e^{i\epsilon_nt}|\phi_n(t)\rangle$. $\epsilon_n$
is quasi-energy and the Floquet state $|\phi_n(t)\rangle$ has the
property $|\phi_n(t)\rangle=|\phi_n(T+t)\rangle$. They are satisfied
the following eigenvalue equation ($\hbar=1$)
\begin{eqnarray} \label{5}
[H(t)-i\frac{\partial}{\partial
t}]|\phi_n(t)\rangle=\epsilon_n|\phi_n(t)\rangle,
\end{eqnarray}
where $H_{eff}=H(t)-i\frac{\partial}{\partial t}$ is defined as the
Floquet Hamiltonian. To solve this equation it is very instructive
to introduce an extend Hilbert space \cite{sambe73} of time-periodic functions with
the inner product
$\langle\langle\cdot|\cdot\rangle\rangle=\frac{1}{T}\int_{0}^{T}dt\langle\cdot|\cdot\rangle$.

With regard to the periodic driving system, it is necessary to make
definite on the time-scales during the evolution. For the case of
Floquet state $|\phi_n(t)\rangle$, since it has the same period with
the driving field, it affects the system dynamics on short
time-scale (in the high-frequency limit). What really affects the
long time-scale of the system dynamics is the gap of the quasi-energies.
Therefore, it is crucial to determine the evolution time through
modulating the structure of quasi-energies in the periodic driving system.

\subsection{Sinusoidal driving}

We first consider the periodic modulation of the  on-site energy for
quantum dot with the sinusoidal form
$\varepsilon(t)=\varepsilon_0\cos(\omega t)$, which can be created
by a waveform generator. In order to obtain an approximate
expression for the quasi-energy, we solve the time-dependent
Schr\"odinger equation by standard perturbation theory
\cite{holthaus92,creffield021}, where the tunneling Hamiltonian is
regarded as the perturbation. Due to the total parity conservation
it is convenient to study in the even parity (or odd parity)
subspace. Then the Hamiltonian reduces $2\times2$ matrix. Since
$H_{\varepsilon}(t)=\varepsilon(t)a^{\dag}a$ is diagonal, when substituting into Eq.(\ref{5}), the
eigenstates of $[H_{\varepsilon}(t)-i\frac{\partial}{\partial t}]$
can be readily given by
\begin{eqnarray}
|\lambda_1(t)\rangle&=&(e^{i\lambda_1t},0)^T,  \nonumber \\
|\lambda_2(t)\rangle&=&(0,e^{i\lambda_2t-i\frac{\varepsilon_0}{\omega}\sin\omega t})^T,
\end{eqnarray}
where $\lambda_{i}$ ($i=1,2$) is the corresponding eigenvalue (i.e., quasi-energy).
On the other
hand, due to the period of Floquet states, one then can find that
the zeroth order approximation of both quasi-energies are zero
(modulo $\omega$). Thus the time-dependent eigenstates can be
approximately viewed as time-independent eigenstates
$|\lambda_1(t)\rangle=(1,0)^T$ and
$|\lambda_2(t)\rangle\simeq(0,1)^T$ in the high-frequency limit
($\omega\gg1$). The first order approximation of quasi-energies can
be obtained via diagonalizing the perturbing matrix
\cite{creffield021}
\begin{eqnarray}
\tilde{H}_t=\left(
    \begin{array}{cc}
      0 & Q \\
      Q^{\ast} & 0 \\
    \end{array}
  \right),
\end{eqnarray}
where the matrix element $Q=\frac{v}{T}\int_{0}^{T}dt
e^{-i\frac{\varepsilon_0}{\omega}\sin\omega t}$ after some straightforward calculations. Consequently, the
quasi-energies are calculated  as $\epsilon_{1,2}=\pm|Q|$ (in the
``first Brillouin zone'') and the corresponding eigenstates become
$|\epsilon_{1,2}\rangle=\frac{1}{\sqrt{2}}(|\lambda_1(t)\rangle\pm|\lambda_2(t)\rangle)$.
The gap of quasi-energies are then given by $\Delta=2 |Q|$. In the
light of the identity
\begin{eqnarray}
e^{i\frac{\varepsilon_0}{\omega}\sin\omega t}=
\sum_{n=-\infty}^{\infty}\mathcal{J}_{n}(\frac{\varepsilon_0}{\omega})e^{i
n\omega t},
\end{eqnarray}
where $\mathcal{J}_{n}$ is the $n$-order Bessel function, we  can
finally obtain the analytical expression for the quasi-energies gap
$\Delta=2|v\mathcal{J}_{0}(\frac{\varepsilon_0}{\omega})|$.

\begin{figure}[h]
\centering
\includegraphics[scale=0.4]{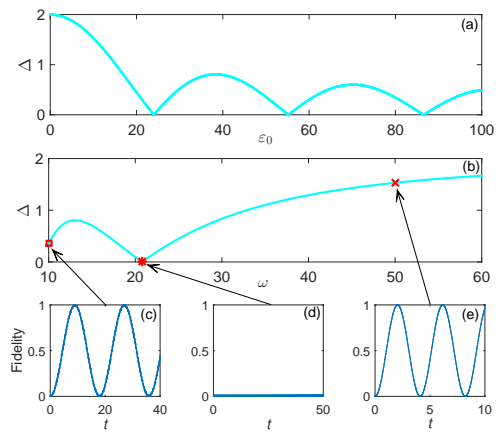}
\caption{The gap of quasi-energies versus (a) the amplitude $\varepsilon_0$ when $\omega=10$, (b) the frequency
$\omega$ when $\varepsilon_0=50$. The gap of quasi-energies approaches
2 when the driving frequency tends to 60 in panel (b). After that the gap
increases slowly with the increasing of driving frequency. The dynamics evolution
of periodic driving system with (c) $\omega=10$, (d) $\omega=20.8$,
(e) $\omega=50$. All parameters are in units of the tunnel coupling
$v$.} \label{fig:03}
\end{figure}

Fig. \ref{fig:03}(a)-(b) demonstrate the relation between the driving
field and the quasi-energies gap, while  Fig. \ref{fig:03}(c)-(e) show that
 the evolution time reduces with the
increasing of the quasi-energies gap. Therefore, we can choose
special evolution time for the periodic driving system by
appropriately selecting the frequency and amplitude of driving
field.
An inspection of Fig. \ref{fig:03}(a) also shows that
the operation time varies with the decreasing of the amplitude of
driving field when we fix a high frequency.
Interestingly, there exists a special case that the
quasi-energies of periodic driving system vanishes (namely the two
quasi-energies approach degeneracy) by choosing the amplitude
$\varepsilon_0$ and the driving frequency $\omega$ properly, which
is known as coherent destruction of tunneling
\cite{creffield02,creffield03}.  As a consequence the state is
localization so that it is invalid to achieve the operation $P_1$,
as shown in Fig. \ref{fig:03}(d).

Since the initial state can be approximately given by
$|\psi(0)\rangle\simeq\frac{1}{\sqrt{2}}(|\epsilon_{1}\rangle-|\epsilon_{2}\rangle)$,
the time evolution of periodic driving system approximately reads
\begin{eqnarray} \label{9a}
|\psi(t)\rangle&\simeq&\frac{1}{\sqrt{2}}(e^{-i\epsilon_{1}t}
|\epsilon_{1}\rangle-e^{-i\epsilon_{2}t}|\epsilon_{2}\rangle) \nonumber\\
&\simeq&\cos{[\mathcal{J}_{0}(\frac{\varepsilon_0}{\omega})vt]}|\lambda_{2}\rangle
- i\sin{[\mathcal{J}_{0}(\frac{\varepsilon_0}{\omega})vt]}|\lambda_{1}\rangle.
\end{eqnarray}
Hence the total operation time for realizing the operation $P_1$ approximately equals
\begin{eqnarray}
\mathcal{T}\simeq\frac{\pi}{2v|\mathcal{J}_{0}(\frac{\varepsilon_0}{\omega})|}.
\end{eqnarray}

\begin{figure}[h]
\centering
\includegraphics[scale=0.35]{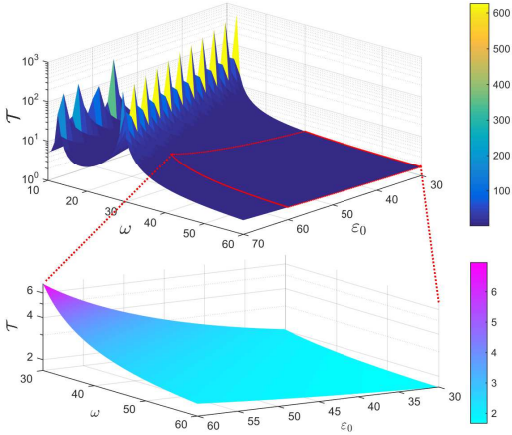}
\caption{The total operation time $\mathcal{T}$ versus  the
amplitude $\varepsilon_0$ and the frequency $\omega$ of the driving
field. } \label{fig:03b}
\end{figure}

\begin{figure}[h]
\centering
\includegraphics[scale=0.45]{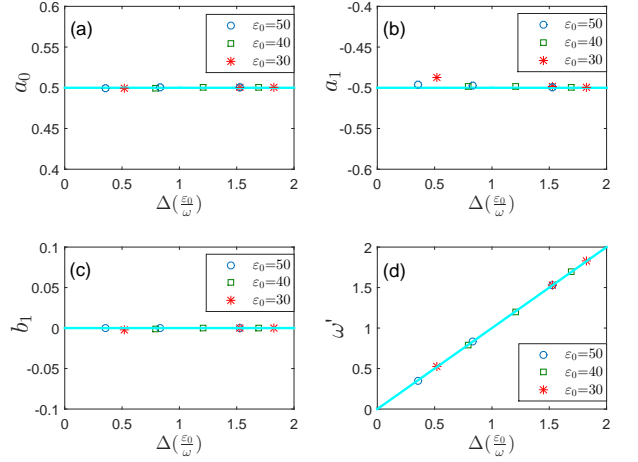}
\caption{The coefficients (a) $a_0$, (b) $a_1$, (c) $b_1$,  (d)
$\omega'$ of fidelity versus the gap of quasi-energies, where the
expression of fidelity for curve-fitting is
$F'=a_0+a_1\cos{\omega't}+b_1\sin{\omega't}$. The lines represent
the analytical solutions given by  Eq. (\ref{11a}) with perturbation
theory while the circles, squares, and stars represent exact results
obtained by curve-fitting. Note that the curve-fitting has high
degree of precision for the exact results since the values of
\emph{R-square} and \emph{Adjusted R-square} approach unit
($\geq99.37\%$) in MATLAB.} \label{fig:031}
\end{figure}

Fig. \ref{fig:03b} depicts the relation between the total  operation
time and the amplitude as well as the frequency of the driving
field. It suggests   that one shall avoid the parameter regions with
the coherent destruction of tunneling, since it takes long operation
time to realize the operation $P_1$. Apart from this regions, the
total operation time can be regulated within proper range. In
addition, from the Eq. (\ref{9a}), one can find readily that the
expression of fidelity is
\begin{eqnarray} \label{11a}
F=\sin^2{[\mathcal{J}_{0}(\frac{\varepsilon_0}{\omega})vt]}
=\frac{1}{2}(1-\cos{[2{\mathcal{J}_{0}(\frac{\varepsilon_0}{\omega})}vt]}).
\end{eqnarray}
Fig. \ref{fig:031} plots the relation between the gap of
quasi-energies and the coefficients of fidelity in the exact and
perturbation regime, respectively. It demonstrates that the
perturbation results work extremely well in the high-frequency
limit.

\begin{figure}[h]
\centering
\includegraphics[scale=0.35]{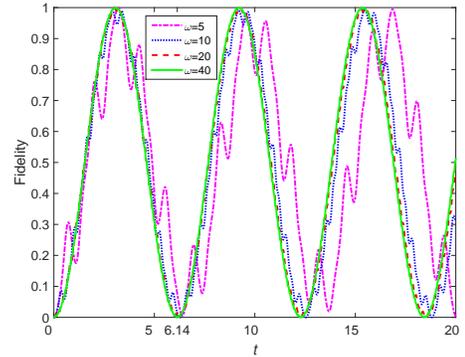}
\caption{The evolution of the fidelity with different  frequencies
of driving field. We have set the quasi-energies gap
$\Delta=2|v\mathcal{J}_{0}(\frac{\varepsilon_0}{\omega})|=1.0236$,
thus the period of system dynamics is approximated to
$T\simeq\frac{2\pi}{\Delta}=6.1383$ in the high-frequency limit,
which is confirmed by the green solid line and red dash line. }
\label{fig:05a}
\end{figure}

In order to check the validity of  the perturbation theory,
we plot the  dynamics of the system with different driving
frequencies. The results are given in Fig. \ref{fig:05a}. We
observe  that the dynamics  is well in  agreement with the results
by perturbation theory  when $\omega>10$ (in units of $v$), while it
deviates seriously from the perturbation results when $\omega<10$ (in
units of $v,$ see the pink dot-dash line in Fig. \ref{fig:05a}).
As a result,  one can employ the perturbation theory safely when the
frequency of the driving field is at least an order of magnitude
larger  than the tunnel coupling.

\begin{figure}[h]
\centering
\includegraphics[scale=0.45]{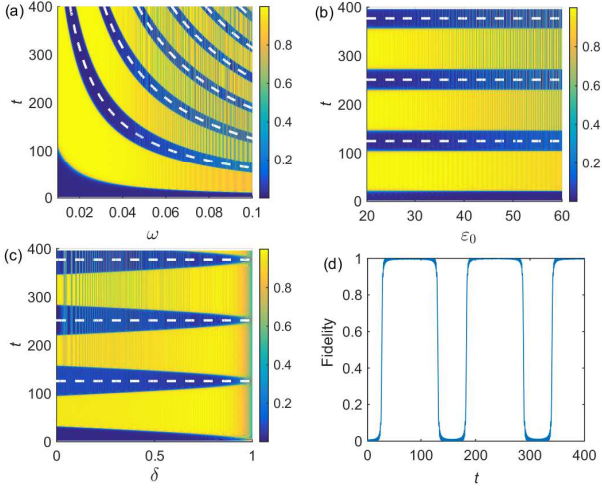}
\caption{The fidelity as a function of the evolution  time and  the
distinct parameters of driving field. The expression of driving
field is $\varepsilon(t)=\varepsilon_0\cos\omega t-\xi$, where $\xi$
is the offset energy of driving field.  (a) $\varepsilon_0=40,
\xi=20$. (b) $\omega =0.05, \xi=0.5\varepsilon_0$. (c)
$\varepsilon_0=40, \omega=0.05, \xi=\delta\varepsilon_0$. (d)
$\varepsilon_0=40, \omega=0.04, \xi=20$. The dash lines are ploted
by the function $t=\frac{2\pi}{\omega}n$, where $n$ is integer.}
\label{fig:05b}
\end{figure}

As mentioned, the perturbation theory is not valid in the
low-frequency limit, this gives rise to a question how the system
behaviors in this limit. Now we go to explore this issue. When the
driving frequency is small such that the
adiabatic condition approximately holds (since the on-site energy changes slowly),
we expected that the system dynamics, e.g., the fidelity, at
the long time-scale would be periodic with
period $T=\frac{2\pi}{\omega}$.  As expectation, we find from the
dash line in Fig. \ref{fig:05b}(a)-(c) that this is exact the
case. Besides, one can observe that  the high fidelity lasts a long
time  within a period when  the driving frequency is small, see Fig.
\ref{fig:05b}(a)-(d). The fidelity changes fast  when the frequency of
driving field is large (see  Fig. \ref{fig:05b}(a),
where the yellow lines and blue lines alter
frequently). It also affects the dynamics when the amplitude of driving field is large,
which is shown in Fig. \ref{fig:05b}(b).
Fig. \ref{fig:05b}(c) illustrates how the offset
energy affects the fidelity. Interestingly, the high fidelity lasts longer time (see the yellow
region) when the offset
on-site energy is larger.

\subsection{Square wave driving}

It is believed that the periodic square wave driving fields are
easily achieved in practice. In fact these driving fields  have been
studied extensively in time-periodic driving system. In particular,
it has been shown in experiment \cite{silveri15} that the
St\"uckelberg interference in a superconducting qubit is driven by
the square wave form which we use in following. We first give the exact
analytical expressions for system evolution operator without any
approximations. The square wave  driving for the on-site energy is
expressed as
\begin{eqnarray}   \label{9}
\varepsilon(t)=\left \{
\begin{array}{rl}
    \varepsilon_1,~~~~~~~~ nT\leq t<t_1+nT,\\
    \varepsilon_2,~t_1+nT\leq t<(n+1)T, \\
\end{array}
\right.
\end{eqnarray}
where $n=1,...,N$ and $t_1\in[0,T]$.
Thus the system evolution operator $U$ of one
period can be written as $U(T,0)=e^{-iH_2t_2}e^{-iH_1t_1}$ with
$t_2=T-t_1$. After some lengthy algebra, one can obtain the expression of the
evolution operator
\begin{eqnarray}
U(T,0)=\frac{1}{x_1x_2}e^{-\frac{i}{2}D}\left(
         \begin{array}{cc}
           A & B+C \\
           B-C & A \\
         \end{array}
       \right),
\end{eqnarray}
where
\begin{eqnarray}
A&=&-4v^2\sin\frac{x_1t_1}{2}\sin\frac{x_2t_2}{2}+(x_1\cos\frac{x_1t_1}{2}+i\varepsilon_1
\sin\frac{x_1t_1}{2})    \nonumber \\
&&\cdot(x_2\cos\frac{x_2t_2}{2}+i\varepsilon_2
\sin\frac{x_2t_2}{2}),    \nonumber \\
B&=&-2iv(x_2\cos\frac{x_2t_2}{2}\sin\frac{x_1t_1}{2}+x_1\cos\frac{x_1t_1}{2}\sin\frac{x_2t_2}{2}),  \nonumber \\
C&=&2v(\varepsilon_1-\varepsilon_2)\sin\frac{x_1t_1}{2}\sin\frac{x_2t_2}{2}, \nonumber \\
D&=&\varepsilon_1t_1+\varepsilon_2t_2,~x_1=\sqrt{\varepsilon_1^2+4v^2},~x_2=\sqrt{\varepsilon_2^2+4v^2}.
\end{eqnarray}

At first we design the driving time $t_1$ ($t_2$) of the on-site
energy $\varepsilon_1$ ($\varepsilon_2$) to satisfy $x_1t_1=\pi$
($x_2t_2=\pi$), that is,
\begin{eqnarray}  \label{12}
t_1=\frac{\pi}{\sqrt{4v^2+\varepsilon_1^2}},~~
t_2=\frac{\pi}{\sqrt{4v^2+\varepsilon_2^2}},~~t_1+t_2=T.
\end{eqnarray}
Consequently, the period of the square wave driving is
confirmed. According to Eq. (\ref{12}), the evolution operator can be further simplified,
\begin{eqnarray}
U(T,0)=\frac{-1}{x_1x_2}e^{-\frac{i}{2}D}\left(
         \begin{array}{cc}
           x_3 & -x_4 \\
           x_4 & x_3 \\
         \end{array}
       \right),
\end{eqnarray}
where $x_3=4v^2+\varepsilon_1\varepsilon_2$ and
$x_4=2v(\varepsilon_1-\varepsilon_2)$.
After $N$ evolution
periods, the final evolution operation becomes
\begin{eqnarray}   \label{14}
U(\mathcal{T},0)&=&U^{N}(T,0)=\frac{1}{2}(\frac{-1}{x_1x_2})^Ne^{-\frac{iND}{2}}   \nonumber\\
    &&\cdot\left(
         \begin{array}{cc}
           r_1^{N}+r_2^{N} & -i(r_1^{N}-r_2^{N}) \\
           i(r_1^{N}-r_2^{N}) & r_1^{N}+r_2^{N} \\
         \end{array}
       \right),
\end{eqnarray}
where $r_1=x_3-ix_4=|r|e^{-i\theta}, r_2=x_3+ix_4=|r|e^{i\theta},
|r|=\sqrt{x_3^2+x_4^2},$ and $\tan\theta=\frac{x_4}{x_3}$. From the
expression in Eq. (\ref{14}), it clearly  requires
$r_1^{N}+r_2^{N}=0$ in order to realize the operation $P_1$
perfectly (up to a global phase factor). By making the vectors
$r_1^{N}$ and $r_2^{N}$ produce a $\pi$-phase difference, that is,
$N\theta-(-N\theta)=\pi$, one can readily obtain the number of
evolution periods
\begin{eqnarray}  \label{15a}
N=\frac{\pi}{2\arctan\frac{2v(\varepsilon_2-\varepsilon_1)}{4v^2+\varepsilon_1\varepsilon_2}}.
\end{eqnarray}

\begin{figure}[h]
\centering
\includegraphics[scale=0.5]{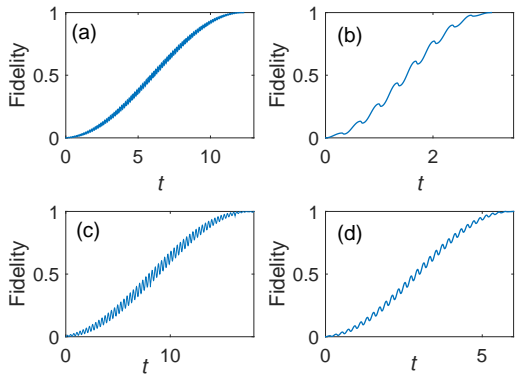}
\caption{The system dynamics of distinct forms for square wave.  (a)
$\varepsilon_1=40$, $\varepsilon_2=60$. (b) $\varepsilon_1=10$,
$\varepsilon_2=90$. (c) $T=0.3$, $N=60$. (d) $T=0.2$, $N=30$. The
other two parameters are calculated by Eq.(\ref{12}) and
Eq.(\ref{15a}).} \label{fig:04}
\end{figure}

Note that according to Eq. (\ref{15a}) the number of evolution periods $N$ is not
integer generally.  Nevertheless it
does not affect the main results because we can just take an integer
nearest to   $N$, as the fidelity increases slowly when it
approaches 1. In turn, if one designates the period $T$ and the
number of evolution periods $N$ in the periodic square wave driving
system, the values of on-site energy $\varepsilon_1$ and
$\varepsilon_2$ can be determined by Eq. (\ref{12}) and Eq.
(\ref{15a}) as well. It demonstrates the system evolution with
distinct values  of $\varepsilon_1$ and $\varepsilon_2$ in Fig.
\ref{fig:04} (a)-(b), as well as the special period and the number of
evolution periods in Fig. \ref{fig:04} (c)-(d). As expected, it can
also realize the operation $P_1$ by square wave driving and we can
modulate the period and the total operation time in this case.

\subsection{ $\delta$-function kick}

When $\varepsilon_2\rightarrow+\infty$, one can readily find
in Eq.(\ref{12}) that $t_2\rightarrow 0$. Then the square wave driving field
reduces  to periodic $\delta$-function kick, i.e.,
\begin{eqnarray}  \label{15}
\varepsilon(t)=\varepsilon_1+\varepsilon_2\sum_{n=1}^{N}\delta(t-nT),
\end{eqnarray}
where $T$ is the driving period and can be calculated approximately as
$T\simeq\frac{\pi}{\sqrt{4v^2+\varepsilon_1^2}}$.
Therefore the total operation time for realizing the operation $P_1$ is approximatively
\begin{eqnarray}
\mathcal{T}&\simeq& \frac{\pi^2}{2\sqrt{4v^2+\varepsilon_1^2}\arctan\frac{2v}{\varepsilon_1}}.
\end{eqnarray}

Note that the dynamics behavior is quite different from  the absence
of $\delta$-function kick, i.e., under a static driving field. In the static case, the evolution operator $U$ reads
\begin{eqnarray}   \label{21}
U(t,0)&=&e^{-\frac{i\varepsilon_{1} t}{2}}\left(
         \begin{array}{cc}
           \cos\frac{x t}{2}+i\frac{\varepsilon_{1}}{x}\sin\frac{x t}{2} & -i\frac{2v}{x}\sin\frac{x t}{2} \\
           -i\frac{2v}{x}\sin\frac{x t}{2} & \cos\frac{x t}{2}-i\frac{\varepsilon_{1}}{x}\sin\frac{x t}{2} \\
         \end{array}
       \right),    \nonumber\\
    x&=&\sqrt{\varepsilon_{1}^2+4v^2}.
\end{eqnarray}
In absence of $\delta$-function kick, the expression of  fidelity
for realizing the operation $P_1$ becomes $|\frac{2v}{x}\sin\frac{x
t}{2}|^2$, where the maximum of fidelity is $|\frac{2v}{x}|^2$. One
easily observes that it cannot obtain the operation $P_1$ when
$\frac{2v}{x}\simeq 0$, i.e., the on-site energy
$\varepsilon_{1}\gg1$. However the situation changes in the
presence of $\delta$-function kick and it can realize the operation
$P_1$ regardless of the large value of on-site energy
$\varepsilon_1$ (the value of $\varepsilon_1$ only determines the driving
period). Especially, when we take away the $\delta$-function kick if
the operation $P_1$ has been completed, the system is still
stationary.

\section{Application to other systems}  \label{V}

\begin{figure}[h]
\centering
\includegraphics[scale=0.5]{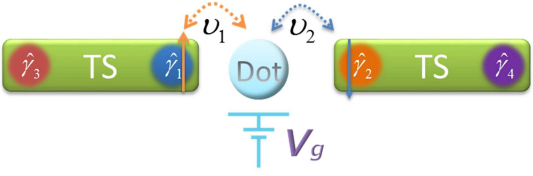}
\caption{The setup for realizing entanglement between the convention qubit and topological qubit.}  \label{fig:05}
\end{figure}

The periodic driving method can be applied to the other structure of
hybrid quantum dot-topological system. Here we apply  it to a
system described by the following Hamiltonian \cite{leijnse11}, as illustrated in Fig. \ref{fig:05},
\begin{eqnarray}
H&=&\varepsilon(t) (\hat{a}_{\uparrow}^{\dag}\hat{a}_{\uparrow}+\hat{a}_{\downarrow}^{\dag}\hat{a}_{\downarrow})
+V\hat{a}_{\uparrow}^{\dag}\hat{a}_{\uparrow}\hat{a}_{\downarrow}^{\dag}\hat{a}_{\downarrow}  +(v_{1}^{\ast}\hat{a}_{\uparrow}^{\dag}-v_{1}\hat{a}_{\uparrow})\hat{\gamma}_1      \nonumber\\
&&+(v_{2}^{\ast}\hat{a}_{\downarrow}^{\dag}-v_{2}\hat{a}_{\downarrow})\hat{\gamma}_2.
\end{eqnarray}
$\varepsilon(t)$ is the on-site energy of the quantum  dot. $v_i
(i=1,2)$ denotes the tunnel coupling between the quantum dot and the
Majorana mode $\hat{\gamma}_i$. In particular, the spin-up (labeled
as $\uparrow$) and spin-down (labeled as $\downarrow$) electrons can
only tunnel into the Majorana mode $\hat{\gamma}_1$ and
$\hat{\gamma}_2$, respectively. $V$ represents the energy
contributed by double occupation on the quantum dot. In the situation  of
large $V$, the quantum dot can only hold single electron.

Since the total parity (the electrons in quantum dot and  the
ordinary fermions formed by Majorana modes) of the hybrid system is
conserved, we can restrict ourself in the even-parity subspace
spanned by $\{|0\rangle_{F_1}|0\rangle_{F_2}|0\rangle_D$,
$|1\rangle_{F_1}|1\rangle_{F_2}|0\rangle_D$,
$|0\rangle_{F_1}|1\rangle_{F_2}|1_{\uparrow}\rangle_D$,
$|0\rangle_{F_1}|1\rangle_{F_2}|1_{\downarrow}\rangle_D$,
$|1\rangle_{F_1}|0\rangle_{F_2}|1_{\uparrow}\rangle_D$,
$|1\rangle_{F_1}|0\rangle_{F_2}|1_{\downarrow}\rangle_D,\}$, where
the subscript $F_i (i=1,2)$ represents the ordinary fermions formed
by Majorana modes. The matrix form of Hamiltonian then can be written as
\begin{eqnarray}
H=\left(
    \begin{array}{cccccc}
      0 & 0 & 0 & v_2 & v_1 & 0 \\
      0 & 0 & v_1 & 0 & 0 & v_2 \\
      0 & v^{\ast}_1 & \varepsilon(t) & 0 & 0 & 0 \\
      v^{\ast}_2 & 0 & 0 & \varepsilon(t) & 0 & 0 \\
      v^{\ast}_1 & 0 & 0 & 0 & \varepsilon(t) & 0 \\
      0 & v^{\ast}_2 & 0 & 0 & 0 & \varepsilon(t) \\
    \end{array}
  \right).
\end{eqnarray}

\begin{figure}[h]
\centering
\includegraphics[scale=0.5]{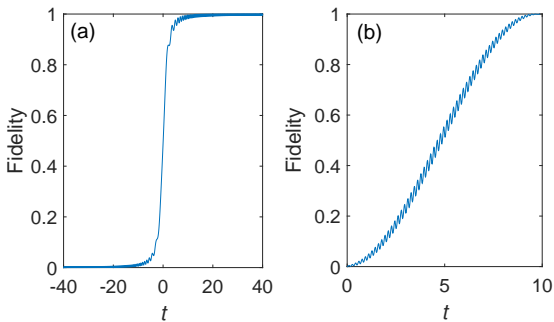}
\caption{ (a) Realizing the operation $P_2$  by the adiabatic
evolution. The on-site energy of the
quantum dot increases with time slowly,
$\varepsilon(t)=1.25t$. (b) Realizing the operation $P_2$  by
 the periodic square-wave driving. $\varepsilon_1=30,
\varepsilon_2=50. $ $\theta=\frac{\pi}{6}$,
$v_1=v_2=\frac{1}{\sqrt{2}}$. All parameters are in units of $v$.}
\label{fig:06}
\end{figure}

This quantum dot-Majorana system \cite{leijnse11,leijnse12}  can be
used to prepare  entanglement between spin and topological qubits or
quantum information transfer between spin and the topological qubits
(even for the quantum logic gates) by the adiabatic  evolution. We
take the preparation of entanglement (denoting as the operation $P_2$)
between the electron spin and Majorana modes as an example to
exemplify how to  manipulate the operation time by periodic square wave driving given in Eq. (\ref{9}).
The operation $P_2$  reads,
\begin{eqnarray}
&P_2&(\sin\theta|0\rangle_{F_1}|0\rangle_{F_2}+ \cos\theta|1\rangle_{F_1}|1\rangle_{F_2})|0\rangle_D  \nonumber\\
&=&\frac{v_1}{v} (\sin\theta|1\rangle_{F_1}|0\rangle_{F_2}+\cos\theta|0\rangle_{F_1}|1\rangle_{F_2})|1_{\uparrow}\rangle_D        \nonumber\\
&&+\frac{v_2}{v} (\sin\theta|0\rangle_{F_1}|1\rangle_{F_2}+\cos\theta|1\rangle_{F_1}|0\rangle_{F_2})|1_{\downarrow}\rangle_D.
\end{eqnarray}
where $v=\sqrt{v_1^2+v_2^2}$. As the Hamiltonian is a $6\times6$ matrix,
the analytical expression of the evolution operator
$U(T,0)=e^{-iH_2t_2}e^{-iH_1t_1}$ is involved. Here we only  give
the equations that determine the period of the driving field and the
total number of evolution periods, i.e.,
\begin{eqnarray}
T&=&t_1+t_2,  \nonumber\\
N&=&\frac{\pi}{2\arctan\frac{x^{\prime}_4}{x^{\prime}_3}},
\end{eqnarray}
where $t_1=\frac{\pi}{\sqrt{\varepsilon_1^2+4(v_1^2+v_2^2)}}$,
$t_2=\frac{\pi}{\sqrt{\varepsilon_2^2+4(v_1^2+v_2^2)}}$,
$x^{\prime}_3=\varepsilon_1\varepsilon_2+4v_1^2+4v_2^2$, and
$x^{\prime}_4=2(\varepsilon_2-\varepsilon_1)\sqrt{v_1^2+v_2^2}$.
Fig. \ref{fig:06} plots the fidelity of realizing operation $P_2$ as
a function of the evolution time by the adiabatic evolution and the
periodic square wave driving, respectively. Again, we find that the
operation time for adiabatic evolution requires relatively long time
since it must   satisfy the adiabatic condition while the operation
time and the period of the square wave driving can be regulated.

\section{discussion and conclusion}  \label{IV}

In the last section, we have studied how to  implement the
operation $P_{1}$ by periodically  driving  the on-site energy of
the quantum dot. For a single operation $P_1$, it is far from
sufficient to   permit quantum computation.  We next briefly discuss
how  to realize an arbitrary rotation for a qubit by successively executing
the operation $P_1$ twice.

As shown in Fig. \ref{fig:01}(b), the system  Hamiltonian of interest
reads
\begin{eqnarray}  \label{26}
H&=&\varepsilon(t) \hat{a}^{\dag}\hat{a}
+(|v_{1}|e^{-i\theta_1}\hat{a}^{\dag}-|v_{1}|e^{i\theta_1}\hat{a})\hat{\gamma}_{1}
\nonumber\\
&&+(|v_{2}|e^{-i\theta_2}\hat{a}^{\dag}-|v_{2}|e^{i\theta_2}\hat{a})\hat{\gamma}_{2},
\end{eqnarray}
where we have introduced  a phase $\theta_1$ ($\theta_2$) into the
tunnel coupling $v_1$ ($v_2$) in the Hamiltonian.  Defining the operator
$\hat{c}=\hat{a}e^{i\theta_1}$, $z=\sqrt{|v_1|^2+|v_2|^2}$,
$z_1=\frac{|v_1|}{z}$, and $z_2=\frac{|v_2|}{z}$, the Hamiltonian (\ref{26})
becomes,
\begin{eqnarray}  \label{27}
H=\varepsilon(t) \hat{c}^{\dag}\hat{c}+z(\hat{c}^{\dag}-\hat{c})(z_1\hat{\gamma}_{1}+z_2\hat{\gamma}_{2}),
\end{eqnarray}
where the phase difference $\theta_1-\theta_2$ equals $2n\pi$ ($n$ is integer and the phase can be modulated by the magnetic flux $\phi$).
The form of Eq.(\ref{27}) is the same as Eq.(\ref{2}) if we redefine
a new Majorana mode
$\hat{\gamma}_{12}=z_1\hat{\gamma}_{1}+z_2\hat{\gamma}_{2}$, where
the tunnel coupling is denoted by $z$. Clearly, the operation
$P_1=\hat{\gamma}_{12}$ in this notation. Consider the two level
system spanned by $\{|0\rangle_F,|1\rangle_F\}$, then we can
express the Majorana operators in terms of Pauli matrices
$\{\sigma_x, \sigma_y, \sigma_z\}$, i.e., $\hat{\gamma}_1=\sigma_x$,
$\hat{\gamma}_2=\sigma_y$, $\hat{\gamma}_1\hat{\gamma}_2=i\sigma_z$.
By successively executing the operation $P_1$ twice with  different
relative tunnel coupling strengthes  between modes $z_1$ and $z_2$,
the total operation becomes
$P=\hat{\gamma}_{12}\hat{\gamma}_{12}'=(z_1\hat{\gamma}_1+
z_2\hat{\gamma}_2)(z_1'\hat{\gamma}_1+z_2'\hat{\gamma}_2)
=(z_1z_1'+z_2z_2')+i(z_1z_2'-z_1'z_2)\sigma_z$,  which is exactly an
arbitrary rotation around the $z$-axis.

Due to the conservation of the total parity, a qubit shall
be encoded by four Majorana modes \cite{bravyi06}. The
Majorana-based qubit can be realized  by the generalization  model in Fig.
\ref{fig:01}(b), which describes a system consisting of  three
quantum dots coupling to four Majorana modes ($\hat{\gamma}_1$,
$\hat{\gamma}_2$, $\hat{\gamma}_3$, $\hat{\gamma}_4$) in the
topological superconductor with comb structure. In the even-parity subspace spanned by
$\{|0\rangle_{F_1}|0\rangle_{F_2}$,
$|1\rangle_{F_1}|1\rangle_{F_2}\}$,  the operation
$P=\hat{\gamma}_{12}\hat{\gamma}_{12}'$ is in fact the rotation
around the $z$-axis, and the operation
$P=\hat{\gamma}_{23}\hat{\gamma}_{23}'$ is the rotation around the
$x$-axis ($\hat{\gamma}_{23}=z_2\hat{\gamma}_2+
z_3\hat{\gamma}_3$), where the ordinary fermion $F_1$ ($F_2$) is formed by the
Majorana modes $\hat{\gamma}_1$ and $\hat{\gamma}_2$
($\hat{\gamma}_3$ and $\hat{\gamma}_4$).

Generally speaking, the tunnel coupling between the quantum
dot and the Majorana mode depends on both the differences among the
on-site energies and the tunnel barriers. By making use of the
periodic driving on  the on-site energy of the quantum dot, the
tunnel coupling would change consequently. Reminding that we can employ
additional electrostatic gates to  manipulate the tunnel barriers,
the tunnel coupling can keep a constant in practise, even the
on-site energies change. Indeed, the possibility of controlling the
tunnel coupling in semiconductor nanowire  has been  experimentally
shown recently \cite{nadjperge10}. So, we believe that in the near
future it is also possible to manipulate  the gates such that the
tunnelling rate remains unchanged in our case, especially with the
periodic square pulses (since it has only two distinct values of on-site energy).

In conclusion, we have proposed a method to regulate the total operation
time of quantum computation, which can be achieved by periodic driving.
By solving the time-dependent
Schr\"odinger equation with perturbation expansion, we have given an
expression of the quasi-energies and elucidated the relationship
between the total operation time and quasi-energies in the
high-frequency limit. As a result, the operation time can be
manipulated  by   designing the amplitude and frequency of the
driving field. For the case of low-frequency limit, due to the
invalidity  of the perturbation theory, we study the dynamical
behaviors by numerical simulations. We find the results approach
those given by adiabatic evolution. Different from the adiabatic
evolution, the system in the low-frequency limit manifests more
intricate   behaviors and the operation time can also be regulated
by the driving field. In particular, the total time  that the high
fidelity lasts  are closely related to the frequency and offset
energy of driving field. For the case of periodic square wave
driving, we have derived an analytical expression for the evolution
operator without any approximations. By this expression, we can calculate the amplitude of
square wave driving fields with fixed operation time and   period of
driving fields. We have also discussed the realization of quantum
operations by the $\delta$-kick, which can be treated as  a deformed
square wave driving. The periodic driving can also be applied  to
the other quantum system---it opens up a new avenue in manipulations
of operation time  in topological quantum computations.

\section*{ACKNOWLEDGMENTS}

We thank S. M. Frolov for helpful discussions.
This work is supported by the National Natural Science Foundation of
China (Grants No. 11175032 and No. 61475033).

\end{document}